\documentclass[universe,article,accept,pdftex,moreauthors]{Definitions/mdpi}

\firstpage{1}
\pubvolume{11}
\issuenum{4}
\articlenumber{119}
\pubyear{2025}
\copyrightyear{2025}
\externaleditor{Máté Csanád} 
\datereceived{17 February 2025}
\daterevised{24 March 2025} 
\dateaccepted{1 April 2025}
\datepublished{4 April 2025}
\hreflink{https://\linebreak doi.org/10.3390/universe11040119} 

\Title{Inclusive Neutrino and Antineutrino Scattering on the $^{12}$C Nucleus Within the Coherent Density Fluctuation Model}

\TitleCitation{Inclusive Neutrino and Antineutrino Scattering on the $^{12}$C Nucleus Within the Coherent Density Fluctuation Model}

\Author{ {Martin} V. Ivanov $^{1,2,}$*\href{https://orcid.org/0000-0001-6855-8279}{\orcidicon} and Anton N. Antonov $^{1}$\href{https://orcid.org/0000-0001-7614-8210}{\orcidicon}}

\AuthorNames{Martin V. Ivanov, Anton N. Antonov}
\isAPAStyle{%
\AuthorCitation{Lastname, F., Lastname, F., \& Lastname, F.}
}{%
\isChicagoStyle{%
\AuthorCitation{Lastname, Firstname, Firstname Lastname, and Firstname Lastname.}
}{
\AuthorCitation{Ivanov, M.V.; Antonov, A.N.}
}
}
\address{%
$^{1}$ \quad Institute for Nuclear Research and Nuclear Energy, Bulgarian Academy of Sciences, Tzarigradsko  {Chaussee} 72, 1784 Sofia, Bulgaria;  {antonovshumen@gmail.com}\\
$^{2}$ \quad Department of Physics, Faculty of Mathematics and Natural Sciences, South-West University ``Neofit Rilski'', 66 Ivan Mihaylov St., 2700 Blagoevgrad, Bulgaria}

\corres{Correspondence:  {martin@inrne.bas.bg}}

\abstract{We investigate quasielastic (anti)neutrino scattering on the $^{12}$C nucleus utilizing a novel scaling variable, $\psi^*$. This variable is derived from the interacting relativistic Fermi gas model, which incorporates both scalar and vector interactions, leading to a relativistic effective mass for the interacting nucleons. For inclusive lepton scattering from nuclei, we develop a new scaling function, denoted as $f^\text{QE}(\psi^*)$, based on the coherent density fluctuation model (CDFM). This model serves as a natural extension of the relativistic Fermi gas (RFG) model applicable to finite nuclei. In this study, we compute theoretical predictions and compare them with experimental data from Miner$\nu$a and T2K for inclusive (anti)neutrino cross-sections. The scaling function is derived within the CDFM framework, employing a relativistic effective mass of $m_N^*$ =  0.8   $m_N$. The findings demonstrate a high degree of consistency with experimental data across all (anti)neutrino energy ranges.}


\begin{document}

\section{Introduction}

The superscaling phenomenon relies on the scaling properties of electron scattering data and was initially considered within the framework of the Relativistic Fermi Gas (RFG) model~\cite{PhysRevC.38.1801, Barbaro1998137, PhysRevLett.82.3212, PhysRevC.60.065502, PhysRevC.65.025502, BCD+04}. At sufficiently high momentum transfer, the inclusive differential $(e,e')$ cross-sections, divided by a suitable function that accounts for the single-nucleon content of the problem, depend on one kinematical variable, known as the scaling $\psi$-variable (this behavior is called scaling of the first kind). When the resulting function is roughly the same for all nuclei, this is referred to as scaling of the second kind. When both kinds of scaling are relevant, the cross-section superscales.

As noted in~\cite{PhysRevC.60.065502}, the true nuclear dynamical content of superscaling is more intricate than that offered by the RFG model. It was observed that the experimental data exhibit a superscaling behavior on the low-$\omega$ side ($\omega$ being the transfer energy) of the quasielastic peak for large negative values of $\psi$ (up to $\psi \approx -2$), while predictions of the RFG model indicate $f(\psi)=0$ for $\psi \leq -1$. This necessitates the consideration of superscaling in realistic finite systems. One approach to this was developed~\cite{A7, A8} within the CDFM~\cite{anton1, anton2, antonov_bjp_1979, antonov_nucleon_1980, antonov_spectral_1982}, which is related to the $\delta$-function limit of the generator coordinate method~\cite{A7, PhysRev.108.311}. It was shown in~\cite{A7, A8} 
that the superscaling in nuclei can be explained quantitatively on the basis of the similar behavior of the high-momentum components of the nucleon momentum distribution in light, medium, and heavy nuclei. It is well known that the latter is related to the effects of the \emph{NN} correlations in nuclei (see,  {e.g.,}~\cite{anton1, anton2}).

{Numerous} studies have been published in recent years to enhance our understanding of lepton--nucleus scattering, as noted in the review report given in Ref.~\cite{Amaro_2020} and references therein. The broad energy distributions associated with neutrino beams result in significant overlap among contributions to the cross-sections, complicating the process of identifying, diagnosing, and correcting the inadequacies of nuclear models. In contrast, electron scattering allows for precise knowledge of energy and momentum transfer. This characteristic enables measurements within defined kinematic ranges and on targets pertinent to neutrino experiments, thereby offering a chance to validate and refine the understanding of nuclear effects. By utilizing electron beams, researchers can probe different interaction mechanisms, measuring the nuclear response at energy transfers that can be varied independently of three-momentum transfer. The study referenced in Ref.~\cite{BENHAR2005230} highlights the significant influence of final-state interactions (FSIs) on the cross-section, particularly in the low $Q^2$ range. In our current research, we do not take FSI into account; since, our results in~\cite{A23} indicate that FSI causes a reduction in the cross-section of less than 4\%, which is largely independent of the neutrino energy when different spectral functions are utilized. In contrast, the effect of meson exchange currents (MECs) on the cross-sections is approximately 25\%, which we included in our calculations.

In our previous studies~\cite{A7, A8, A10}, 
we obtained the CDFM scaling function $f(\psi)$ using the scaling function obtained within the RFG model and averaging it with the weight function $|F(x)|^2$, which is equivalently associated with either the density $\rho(r)$ or the nucleon momentum distribution $n(k)$ in nuclei. Consequently, the CDFM scaling function consists of an infinite superposition of weighted RFG scaling functions. This method enhances the RFG model and the characterization of the scaling function for realistic finite nuclear systems. In our work~\cite{A10}, we managed to describe experimental data of the inclusive electron scattering in the QE-region utilizing the CDFM function, derived from parameterizing the RFG scaling function with the coefficient $c_1$, which addresses the observed asymmetry of the scaling function. The coefficient value $c_1$ ($c_1 \neq 3/4$) is chosen based on empirical data, relating to the momentum transfer at the peak. Additionally, cross-sections for a variety of processes have been calculated using the CDFM scaling function, including inclusive electron scattering in the QE and $\Delta$-regions
, along with neutrino (antineutrino) scattering for both charge-changing (CC) 
and neutral-current (NC) 
processes.

In a recently published study~\cite{PhysRevC.109.064621}, we established and explored a novel scaling approach known as CDFM$_{M^*}$, which incorporates an effective mass $m_N^*$ and defines $M^* = m_N^*/m_N$. This approach utilizes the scaling function derived from the CDFM, which is fundamentally based on the scaling function of the RFG. Additionally, we used a new scaling variable $\psi^*$, which is derived from the scaling characteristics of the Relativistic Mean Field (RMF) model in nuclear matter~\cite{Ama15}. As shown in Refs.~\cite{Ama15, Ama17, Mar17, Ama18, Rui18}, the newly defined scaling function $f^*(\psi^*)$ integrates dynamical relativistic effects through the incorporation of an effective mass. Within the framework of the CDFM$_{M^*}$ model, we derived a scaling function $f^\text{QE}(\psi^*)$ applicable in the quasielastic (QE) region, utilizing the empirical density distribution of protons to ascertain the weight function $|F(x)|^{2}$. Using the scaling function $f^\text{QE}(\psi^*)$, we computed the longitudinal and transverse response functions through two methodologies: one employing the scaling function from CDFM$_{M^*}$ and the other using the conventional scaling function from the CDFM model (where ${M^*=1}$). The results indicate that the CDFM$_{M^*}$ model enhances the transverse components of the electromagnetic current. This observation supports the conclusion that an effective nucleon mass reduction to $M^* = 0.8$ results in an increased transverse response, consistent with the RMF model's inclusion of dynamical relativistic effects, such as the enhancement of the transverse response attributed to the lower components of nucleon spinors. Furthermore, we calculated the inclusive $(e,e')$ and (anti)neutrino differential quasielastic cross-section utilizing the scaling function $f^\text{QE}(\psi^*)$. This analysis incorporated the 2p-2h MEC contribution derived from the RFG model. The findings indicate a satisfactory representation of the electron data~\cite{RevModPhys.80.189} and the (anti)neutrino data from MiniBooNE~\cite{PhysRevD.81.092005, miniboone-ant}.

In the present work, we use the CDFM$_{M^*}$ approach to calculate (anti)neutrino CC quasielastic differential cross-sections and compare results with the data from T2K and Minerva experiment. The model also includes the contribution of weak two-body currents in the two-particle two-hole sector, evaluated within a fully relativistic Fermi gas. Section~\ref{sec:scheme} contains a brief review of the theoretical scheme used to obtain the CDFM$_{M^*}$ scaling function and to apply it to calculations of the (anti)neutrino CC quasielastic differential cross-sections. In Section~\ref{sec:results}, we present our main results for the latter. Finally, in Section~\ref{sec:conclusions}, we draw the main conclusions of the present work.

\section{Theoretical Scheme\label{sec:scheme}}

In this section, we give briefly the theoretical scheme used to calculate the $(\nu_\mu,\mu^-)$ and $(\overline{\nu}_\mu,\mu^+)$ cross-sections within the CDFM$_{M^*}$ model. Let us denote the energy of the incident neutrino as $\epsilon=E_\nu$ and of the detected muon as $\epsilon'=m_\mu+T_\mu$ with momenta  {${\bf k}$}  and ${\bf k}'$, respectively.

The expressions for the  four-momentum transfer are \mbox{$k^\mu-k'{}^\mu=(\omega,{\bf q})$} and $Q^2={\bf q}^2-\omega^2 > 0$, with ${\bf q}$ and $\omega$ being the three momentum and energy transfer, respectively. {For the (anti)neutrino the scattering angle $\theta_\mu$,  the double-differential cross-section is the following~\cite{Ama05a, Ama05b}:
\begin{equation}
\left[\frac{d^2\sigma}{dT_\mu d\cos\theta_\mu}\right]_\chi=\sigma_0{\cal F}_\chi^2
\, ,\label{nuxsec0}
\end{equation}
where $\chi = +$ for neutrino-induced reactions (for example, $\nu_l +n \rightarrow l^- +p$, where \mbox{$l=e,\mu, \tau$) and $\chi=-$} for antineutrino-induced reactions (for example, $\overline{\nu_l} +p \rightarrow l^+ +n$, where $l=e,\mu, \tau$). In Equation~(\ref{nuxsec0}),
\begin{equation}
\sigma_0=
\frac{G^2\cos^2\theta_c}{4\pi}
\frac{k'}{\epsilon}v_0,
\end{equation}
where $G=1.166\times 10^{-11}~\rm MeV^{-2} \sim 10^{-5}/ m_p^2$ is the Fermi constant, $\theta_c$ is the Cabibbo angle, $\cos\theta_c=0.975$, and $v_0= (\epsilon+\epsilon')^2-q^2$ is the kinematic factor. The nuclear-structure-dependent quantity ${\cal F}_\chi^2$ may be written as
\begin{equation}
{\cal F}_\chi^2= [V_{CC} R_{CC} + 2{V}_{CL} R_{CL} + {V}_{LL} R_{LL} + {V}_{T} R_{T}] \pm \chi [2{V}_{T'} R_{T'}] \, ,\label{nuxsec}
\end{equation}
that is, as a generalized Rosenbluth decomposition having charge--charge ($CC$), charge--longitudinal ($CL$), longitudinal--longitudinal ($LL$), and two types of transverse ($T,T'$) responses. In Equation~(\ref{nuxsec}), $V_K$ are the lepton kinematics factors and $R_K(q,\omega)$ ($K=CC$, $CL$, $LL$, $T$, and $T'$) are the nuclear response functions. The function $R_{T'}$ is added with ($+$) for neutrinos and subtracted ($-$) for antineutrinos. At this point, we note that the nuclear response functions $R_K$  are proportional to the single-nucleon response function $U_K$ multiplied by the scaling function $f^\text{QE}(\psi^*)$
\begin{equation} \label{cdfmm*}
R_K = \frac{{\cal N} \xi^*_F}{m^*_N \eta_F^{*3} \kappa^*}  U_K  f^\text{QE}(\psi^*).
\end{equation}

{The} expressions of the single nucleon response functions $U_K$ and the lepton kinematics factors $V_K$ ($K=CC$, $CL$, $LL$, $T$, and $T'$) are given in Ref.~\cite{Rui18}.} In Equation~(\ref{cdfmm*}), $\psi^*$ is the scaling variable

\begin{equation}
\psi^* = \sqrt{\frac{\epsilon_0^*-1}{\epsilon_F^*-1}} {\rm sgn} (\lambda^*-\tau^*),\label{psi*}
\end{equation}
which is the minimum kinetic energy of the initial nucleon divided by the kinetic Fermi energy. Other quantities are as follows (see also~\cite{PhysRevC.109.064621}):
\begin{eqnarray}
\xi_F^*  &=&  \sqrt{1+{\eta_F^*}^2}-1,\\
\eta_F^*  &=&   k_F/m_N^*,\\
\kappa^*   &=&  q/(2m_N^*),\\
\lambda^*  &=& \omega/(2m_N^*),\\
\tau^*  &=&  {\kappa^*}^2-{\lambda^*}^2, \\
\epsilon_F^* &=& \sqrt{1+{\eta_F^*}^2},\\
\epsilon_0^*&=&{\max}
\left\{
\kappa^*\sqrt{1+\frac{1}{\tau^*}}-\lambda^*, \epsilon_F^*-2\lambda^*\label{eps0*}
\right\}
\end{eqnarray}
and $m^*_N$ is the nucleon effective mass. For the free Dirac and Pauli form factors, we use the Galster parametrization~\cite{GALSTER1971221}.

{Here, we briefly provide some basic relationships from the ground of the CDFM \mbox{model~\cite{anton1, anton2, antonov_bjp_1979, antonov_nucleon_1980, antonov_spectral_1982}.} The model is based on the generator coordinate method (GCM) which has been suggested by Hill and Wheeler in~\cite{PhysRev.89.1102} and Griffin and Wheeler in~\cite{PhysRev.108.311}. In it, the trial many-body wave function $\Psi({\{\boldsymbol{r}}_i\})$ of the system of $A$ nucleons is given in the form of a \mbox{linear combination}:
\begin{equation}
\Psi(\boldsymbol{r}_1,\dots,\boldsymbol{r}_A)=\int F(x_1,x_2,\dots)\Phi(\boldsymbol{r}_1,\dots,\boldsymbol{r}_A;x_1,x_2,\dots)\mathrm{d}x_1\mathrm{d}x_2\dots,
\end{equation}
where $\Phi(\boldsymbol{r}_1,\dots,\boldsymbol{r}_A;x_1,x_2,\dots)$ is the generating function (usually chosen to be a Slater determinant) which depends on the radius-vectors of the particles $\{\boldsymbol{r}_i\}$ and on the generator coordinates $x_1,x_2,\dots$. The ``weight'' function $F$ is determined using the \mbox{variational principle}
\vspace{-6pt}\begin{equation}
\delta E=0,\label{2a}
\end{equation}
where
\begin{equation}
E[\Psi]=\langle\Psi|\hat{H}|\Psi\rangle/\langle\Psi|\Psi\rangle,
\end{equation}
{$\hat{H}$} being the Hamiltonian of the system. Equation~(\ref{2a}) leads to the integral equation for the weight function
\vspace{6pt}
\begin{equation}
\int [\mathcal{H}(x,x')-E\mathcal{I}(x,x')]f(x')\mathrm{d}x'=0,\label{4a}
\end{equation}
where the energy kernel $\mathcal{H}(x,x')$ and the overlap kernel $\mathcal{I}(x,x')$ have the forms
\begin{gather}
\mathcal{H}(x,x') = \langle\Phi(\left\{\boldsymbol{r}_i\right\},x)|\hat{H}|\Phi(\left\{\boldsymbol{r}_i\right\},x')\rangle\label{5a}\\
\mathcal{I}(x,x') = \langle\Phi(\left\{\boldsymbol{r}_i\right\},x)|\Phi(\left\{\boldsymbol{r}_i\right\},x')\rangle,\label{6a}
\end{gather}
where the generator coordinate $x$ denotes $x_1, x_2, \dots$.

In the case of a many-fermion system, the kernels $\mathcal{H}$ and $\mathcal{I}$ peak strongly at $x\sim x'$, which leads to the extreme case of the delta-function limit of GCM. The latter consists in
\begin{align}
& \mathcal{I}\left(x, x^{\prime}\right) \rightarrow \delta\left(x-x^{\prime}\right),\label{7a} \\
& \mathcal{H}\left(x, x^{\prime}\right) \rightarrow-\frac{\hbar^2}{2 m_{\mathrm{eff}}} \delta^{\prime \prime}\left(x-x^{\prime}\right)+\delta\left(x-x^{\prime}\right) V\left(\frac{x+x^{\prime}}{2}\right).\label{8a}
\end{align}

{The} substitutions~(\ref{7a}) and~(\ref{8a}) lead the integral Equation~(\ref{4a}) to a Shr\"{o}dinger-type equation (this point being discussed by Dirac in Ref.~\cite{Dirac_1930} and by Griffin and Wheeler in Ref.~\cite{PhysRev.108.311}).

It was shown within the basic considerations of the CDFM (see~\cite{anton1, anton2, antonov_bjp_1979, antonov_nucleon_1980, antonov_spectral_1982}) that the one-body density matrix can be written as a coherent superposition of one-body density matrices for spherical ``pieces'' of radius $x$ (the so-called ``fluctons'') containing $A$ nucleons homogeneously distributed with density
\begin{equation}
\rho_{0}(x)=\frac{3A}{4\pi x^{3}}.\label{9a}
\end{equation}

{The} CDFM one body density matrix has the following form:
\begin{equation}
\rho\left(\boldsymbol{r}, \boldsymbol{r}^{\prime}\right)=\int_0^{\infty}|F(x)|^2 \rho_x\left(\boldsymbol{r}, \boldsymbol{r}^{\prime}\right) \mathrm{d} x
\end{equation}
with
\begin{equation}
\rho_x\left(\boldsymbol{r}, \boldsymbol{r}^{\prime}\right)=3 \rho_0(x) \frac{j_1\left(k_{\mathrm{F}}(x)\left|\boldsymbol{r}-\boldsymbol{r}^{\prime}\right|\right)}{k_{\mathrm{F}}(x)\left|\boldsymbol{r}-\boldsymbol{r}^{\prime}\right|} \theta\left(x-\frac{\left|\boldsymbol{r}+\boldsymbol{r}^{\prime}\right|}{2}\right) .\label{11a}
\end{equation}

{In} Equation~(\ref{11a}), $j_1$ is the first-order spherical Bessel function and
\begin{equation}
\theta (y) = \left\{\begin{array}{c}
1,\quad y\geq0 \\
0,\quad y < 0
\end{array}
\right. \label{12a}
\end{equation}
is the step-function of Heaviside.
}

The scaling function $f_\text{RFG}^\text{QE}[\psi^*(x)]$ with
\begin{equation}
\psi^*(x)=\dfrac{k_{F}x\psi^*}{\alpha}\label{eq:5}
\end{equation}
and
\begin{equation}
\alpha=(9\pi A/8)^{1/3}\simeq 1.52A^{1/3}
\end{equation}
is related to the RFG model (for a system of a ``flucton'' with radius $x$) and has the following form:
\begin{equation}
f_\text{RFG}^\text{QE}[\psi^*(x)] = \displaystyle \frac{3}{4} \left[\! 1\!-\!\left( \frac{k_F x \psi^*}{\alpha} \right)^{2}\!\right]
\theta \left(\!1-\!\left(\dfrac{k_F x\psi^*}{\alpha}\right)^2\!\right). \label{eq:4}
\end{equation}

As shown in Ref.~\cite{PhysRevC.109.064621} in the CDFM$_{M^*}$, the scaling function $f_\text{RFG}^\text{QE}[\psi^*(x)]$ is averaged by the weight function $|F(x)|^2$ and the scaling function of the nucleus becomes the following:
\begin{equation}
f^\text{QE}(\psi^*)= \int_{0}\limits^{\infty}  |F(x)|^{2} \,f_\text{RFG}^\text{QE}[\psi^*(x)]\,{\mathop{}\!\mathrm{d}} x. \label{eq:1}
\end{equation}

{In} Equation~(\ref{eq:1}),
\begin{equation}
|F(x)|^{2}=-\frac{1}{\rho_{0}(x)} \left. \frac{\mathrm{d}\rho(r)}{\mathrm{d}r}\right |_{r=x}, \label{eq:2}
\end{equation}
with $\rho(r)$ being the nucleon density distribution in the case of a monotonically decreasing one ($\mathrm{d}\rho(r)/\mathrm{d}r \leq0$). {The densities $\rho(r)$ (in fm$^{-3}$) of $^{56}$Ni and $^{208}$Pb calculated in the Skyrme \mbox{HF + BCS} method with SLy4 force (normalized to $A=56$ and $A=208$, respectively) and the weight function $|F(x)|^2$ (in fm$^{-1}$) normalized to unity are shown in Figure~1 of Ref.~\cite{astronomy2010001}. It is clearly visible that the primary contribution to the weight functions is derived from the surface region of the density distribution, where $\rho(r)$ decreases monotonically. The central density region, which may contain some fluctuations in $\rho(r)$, accounts for less than 1\% of the normalization of the weight function.}

It was shown in~\cite{A8} (see also~\cite{PhysRevC.109.064621}) that the integration in Equation~(\ref{eq:1}), using \mbox{Equation~(\ref{eq:2}),} leads to the following expression for the scaling function:
\begin{equation}
f^\text{QE}(\psi^*)= \frac{4\pi}{A}\int\limits_{0}^{\alpha/(k_{F}|\psi^*|)} \rho(x) \Bigg[ x^2 f_\text{RFG}^\text{QE}[\psi^*(x)]+ \frac{x^3}{3} \frac{\partial f_\text{RFG}^\text{QE}[\psi^*(x)]}{\partial x} \Bigg]\, {\mathop{}\!\mathrm{d}} x. \label{eq:8}
\end{equation}

{We} note that in Equations~(\ref{eq:5}), (\ref{eq:4}), and~(\ref{eq:8}), the Fermi momentum $k_F$ is not a free parameter (as it is in the RFG model) and can be calculated in the CDFM$_{M^*}$ model for a given nucleus using the following relationship:
\begin{equation}
k_F= \int\limits_{0}^{\infty} |F(x)|^2 \,k_{x}(x)\, {\mathop{}\!\mathrm{d}} x=  \int\limits_{0}^{\infty}  |F(x)|^{2}\,\frac{\alpha}{x}\, {\mathop{}\!\mathrm{d}} x= \frac{4\pi(9\pi)^{1/3}}{3A^{2/3}} \int\limits_{0}^{\infty} \rho(r)\, r\, {\mathop{}\!\mathrm{d}} r \label{eq:6}
\end{equation}
under the condition
\begin{equation}
\lim_{r\rightarrow \infty} \left[ \rho(r)\,r^2 \right]=0. \label{eq:7}
\end{equation}

{At} this point, we note that in the present work, the weight and the scaling functions in the CDFM$_{M^*}$ model are normalized as follows:
\begin{equation}\label{norm}
\int\limits_{0}^{\infty} |F(x)|^2 {\mathop{}\!\mathrm{d}} x =1,\quad
\int\limits_{-\infty}^{\infty}f^\text{QE}(\psi^*){\mathop{}\!\mathrm{d}} \psi^* = 1.
\end{equation}

\section{Results and {Discussion}\label{sec:results}}

{In}  this section, we present our main results for the inclusive (anti)neutrino CC quasielastic differential cross-sections using the new scaling function of the CDFM$_{M^*}$ model. A test of the CDFM$_{M^*}$ model for inclusive ($e,e'$) scattering is provided in our \mbox{paper~cited in \cite{PhysRevC.109.064621}.} We demonstrated that the CDFM$_{M^*}$ model description of inclusive ($e,e'$) scattering data is quite acceptable using just one parameter, namely, the effective mass $M^*$, which is fixed to $0.8$ in all performed calculations. Also, in Ref.~\cite{PhysRevC.109.064621}, the CDFM$_{M^*}$ model is applied to CCQE (anti)neutrino scattering data of the MiniBooNE experiment. We found that the scaling approach CDFM$_{M^*}$, which includes both QE and 2p-2h MEC, leads to results that are in good agreement with (anti)neutrino scattering data.

\begin{figure}[H]
\begin{adjustwidth}{-\extralength}{0cm}
\includegraphics[width=0.99\linewidth]{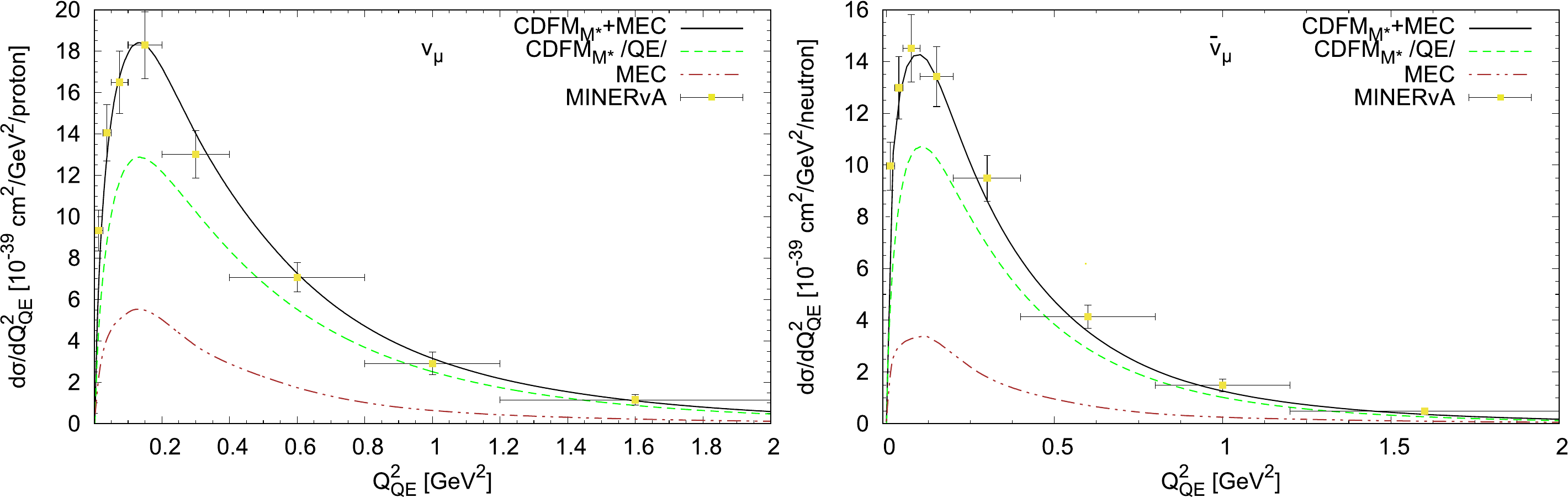}
\end{adjustwidth}
\caption{ {Flux}$-$folded CCQE $\nu_\mu{-}^{12}$C (left panel) and $\overline{\nu}_\mu{-}^{12}$C (right panel) scattering cross-section per target nucleon as function of $Q^2_\text{QE}$ obtained within CDFM$_{M^*}$ model including MEC. 2p$-$2h MEC and QE results are shown separately. MINER$\nu$A data are from~\cite{PhysRevLett.111.022502, PhysRevLett.111.022501}.\label{fig1}}

\end{figure}

In the present paper, we extend the theoretical analysis of neutrino and antineutrino CCQE scattering comparing our results with the data of two other experiments. The results in Figure~\ref{fig1} correspond to the MINER$\nu$A flux-averaged CCQE $\nu_\mu$($\overline{\nu}_\mu$) differential cross-section per nucleon as a function of the reconstructed four-momentum $Q^2_\text{QE}$. The latter is obtained in the same way as for the experiment, assuming an initial state nucleon at rest with a constant binding energy, $E_b$, set to $34$~MeV ($30$~MeV) in the neutrino (antineutrino) case. The left panel refers to $\nu_\mu$--$^{12}$C, whereas the right panel contains predictions and data for $\overline{\nu}_\mu$--CH. The results, which include both QE derived from the CDFM$_{M^*}$ model and the 2p-2h MEC contributions, are compared with the experimental data presented in Figure~\ref{fig1}. The figures also separately illustrate the contributions from QE and 2p-2h MEC. It is noteworthy that the 2p-2h MEC plays a crucial role in accurately describing the experimental data, contributing roughly 20--30\% to the total response at its maximum. The theoretical predictions from the CDFM$_{M^*}$ model, which encompass both QE and 2p-2h MEC contributions, align very well with the observed data.\newpage

In this work, we utilize the 2p-2h MEC model developed in Ref.~\cite{Simo:2016ikv}, which is an extension to the weak sector of the seminal papers~\cite{VanOrden:1980tg, DEPACE2003303, Amaro:2010iu} for the electromagnetic case. The calculation is entirely based on the RFG model with an effective mass of $M^*=1$. A general parametrization of the MEC responses is employed, which considerably decreases computational time. For the specific cases of $^{12}$C and $^{16}$O, this parametrization is detailed in Refs.~\cite{PhysRevD.94.093004, PhysRevD.94.013012, Megias2018}. It is important to highlight that an alternative parametrization for electroweak 2p2h MEC responses, derived from the RMF with an effective mass of $M^*=0.8$, has been presented in Refs.~\cite{PhysRevC.104.025501, PhysRevD.104.113006}. This new approach may offer advantages compared to the CDFM$_{M^*}$ model. Consequently, the first step in improving our model is to implement this alternative parametrization to assess its potential impact on future investigations.

\begin{figure}[H]
\begin{adjustwidth}{-\extralength}{0cm}
\includegraphics[width=0.99\linewidth]{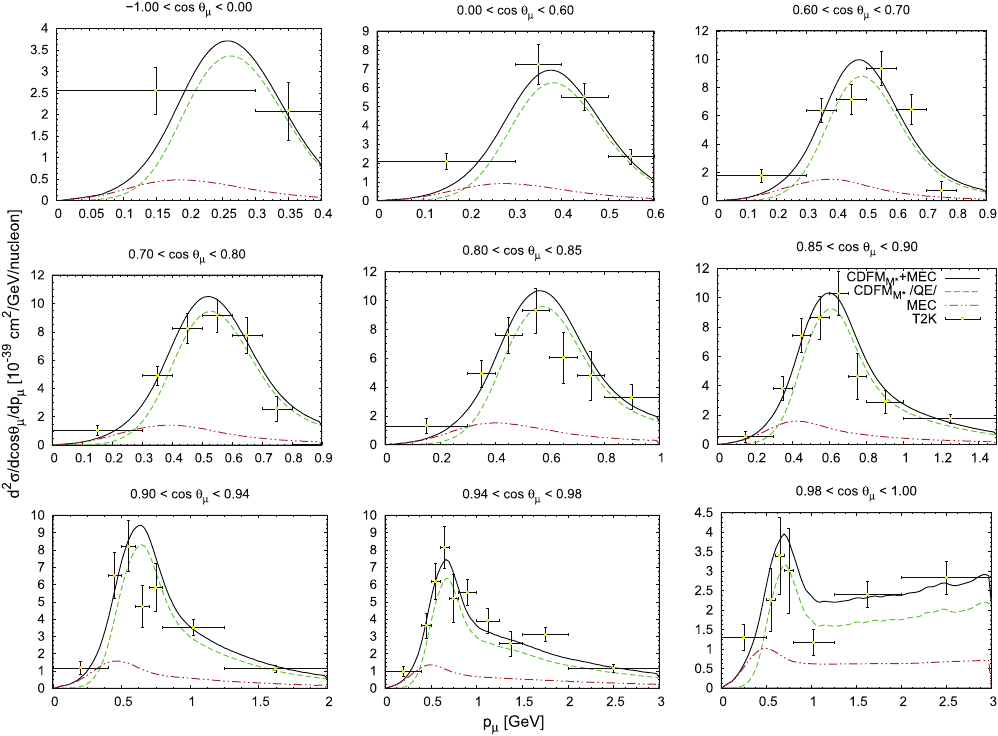}
\end{adjustwidth}
\caption{ {T2K} flux$-$folded double differential cross$-$section per target nucleon for the $\nu_\mu$ CCQE process on $^{12}$C displayed versus the $\mu^{-}$ momentum $p_\mu$ for various bins of $\cos \theta_\mu$ obtained within the CDFM$_{M^*}$ model including MEC. 2p$-$2h MEC and QE results are shown separately. The data are from Ref.~\cite{PhysRevD.93.112012}.\label{fig2}}

\end{figure}

In Figure~\ref{fig2}, we present the flux-averaged double differential cross-sections corresponding to the T2K experiment~\cite{PhysRevD.93.112012}. The graphs are plotted against the muon momentum, and each panel corresponds to a bin in the scattering angle. As in the previous case, we show the results obtained within the CDFM$_{M^*}$ model including MEC and also the separate QE and 2p-2h MEC contributions. As already pointed out in~\cite{PhysRevD.94.093004}, the narrower T2K flux, sharply peaked at about $0.7$~GeV, is the reason for the smaller contribution provided by the 2p-2h MEC (of the order of $\sim$10$\%$, only at very forward angles the contribution of the 2p-2h MEC is a larger up to 25\% in the maximum of the QE peak) as compared with the MINER$\nu$A results: in fact, the main contribution for the 2p-2h response comes from momentum transfers $q\sim$500 MeV/c, which are less important at T2K kinematics. Regarding the theoretical predictions, the CDFM$_{M^*}$ model shows very good agreement with the experimental T2K data for all considered angles.

\section{Conclusions}\label{sec:conclusions}

The considerations and the results of the present work can be summarized as follows:

\begin{itemize}

\item[(i)] The scaling function $f^\text{QE}(\psi^*)$, which has been constructed within the Coherent Density Fluctuation Model (CDFM$_{M^*}$) using the new scaling variable $\psi^*$ from the interacting relativistic Fermi gas model with scalar and vector interactions, is used for analyses of (anti)neutrino scattering processes. The interacting relativistic Fermi gas model is known to generate a relativistic effective mass for the interacting nucleons. The CDFM$_{M^*}$ model itself is based on the $\delta$-function limit of the generator coordinate method and is a natural extension of the relativistic Fermi gas model to finite nuclei. It keeps the gauge invariance and describes the dynamical enhancement of both the lower components of the relativistic spinors and transverse response function. We should emphasize that it makes it possible to explain quantitatively the superscaling in realistic finite systems on the basis of the similar behaviour of the high-momentum components of the nucleon momentum distribution from light to heavy nuclei. As is known, the latter is a result of the effects of \emph{NN} correlations in nuclei.

\item[(ii)] The scaling approach of CDFM$_{M^*}$ is used to calculate inclusive (anti)neutrino charge-changing (CC) quasielastic (QE) differential cross-sections and to compare the theoretical results with the experimental data from MINER$\nu$A~\cite{PhysRevLett.111.022502, PhysRevLett.111.022501} and T2K~\cite{PhysRevD.93.112012} experiments in addition to our previous consideration~\cite{PhysRevC.109.064621} of the data from the \mbox{MiniBooNE~\cite{PhysRevD.81.092005, miniboone-ant}} and NOMAD~\cite{Lyubushkin:2009} experiments. In the calculations, a value of relativistic effective mass $m^*_N= 0.8 m_N$ is used. The only free parameter present in our model is the latter one. In Ref.~\cite{Ama15}, the authors explore the $\psi^*$  scaling idea in the context of the RMF for nuclear matter. In the study, the best value of the effective mass $M^*=m^*_N/m_N=0.8$ is obtained.  {This value provides the best scaling behavior of the data with a large fraction of data concentrated around the universal scaling function of the RFG model while the CDFM model is its natural extension to finite nuclei.}

We note also that the Fermi momentum $k_F$ is not a free parameter and is calculated in the CDFM$_{M^*}$ model for a given nucleus. In the present work, we used the 2p-2h MEC model developed in Ref.~\cite{Simo:2016ikv} with a general parametrization of the MEC responses.

\item[(iii)] The results of the flux-folded CCQE $\nu_\mu$--$^{12}$C and $\overline{\nu}_\mu$--CH scattering cross-section per target nucleon as a function of $Q_\text{QE}^2$ including MEC are presented and compared with the data from the MINER$\nu$A \cite{PhysRevLett.111.022502, PhysRevLett.111.022501} in Figure~\ref{fig1}. The results of our calculations for the $\nu_\mu$ CCQE process on $^{12}$C versus the $\mu^-$ momentum $p_{\mu}$ for various bins of $\theta_\mu$ including MEC are given in Figure~\ref{fig2} and compared with the data from the T2K experiment Ref.~\cite{PhysRevD.81.092005, miniboone-ant}. The theoretical results obtained within the CDFM$_{M^*}$ model, including both QE and 2p-2h MEC, are in very good agreement with the data from the considered MINER$\nu$A and T2K experiments in most of the kinematical situations considered in the present work. It is shown that the contribution of the 2p-2h MEC effects can be of the order of $\sim$20--30\% compared with the pure QE responses in the case of Miner$\nu$a and $\sim$10\% in the T2K kinematics (only for very forward angles up to 25\%). The results of the present work allow us to conclude that our approach is capable of being applied to analyses of (anti)neutrino scattering on the $^{12}$C nucleus. {Our choice of $^{12}$C is mainly due to the available experimental data of neutrino (antineutrino) scattering on this nucleus. Here, it is important to emphasize that our approach is easily extendable to heavier nuclei, as the foundation of the superscaling phenomenon is that the scaling function is the same for all nuclei. The latter is due to the similar behaviour of the high-momentum tail of the nucleon momentum distribution, being effects of nucleon-nucleon correlations. Our previous studies have established that the scaling function obtained within the CDFM model demonstrates a \mbox{superscaling behavior}.}

{
\item[(iv)] As a general conclusion, we note that in the present work, we consider the important problem of the neutrino--nuclei interaction, which is related to the fundamental question of the neutrino oscillations. Our approach uses the CDFM$_{M^*}$ model, which is gauge-invariant, and allows one to transform the nuclear matter quantities to the corresponding ones in finite nuclei. The model includes the possibility of accounting for the nucleon--nucleon correlations in nuclei whose effects are responsible for the superscaling, which is one of the most important phenomena when it comes to lepton interactions with nuclei.

}
\end{itemize}


\authorcontributions{Conceptualization, M.V.I. and A.N.A.; methodology, M.V.I. and A.N.A.; software, M.V.I.; validation, M.V.I. and A.N.A.; writing---original draft preparation, M.V.I. and A.N.A.; writing---review and editing, M.V.I. and A.N.A. All authors have read and agreed to the published version of the manuscript.}

\funding{This research received no external funding.}

\dataavailability{The original contributions presented in the study are included in the article; further inquiries can be directed to the corresponding author(s).}

\conflictsofinterest{The authors declare no conflicts of interest.}

\begin{adjustwidth}{-\extralength}{0cm}

\reftitle{References}

%
\PublishersNote{}
\end{adjustwidth}

\end{document}